\documentclass[aps,prb,reprint,groupedaddress]{revtex4-1}

\usepackage{graphicx} % Include figure files
\usepackage{dcolumn}  % Align table columns on decimal point
\usepackage{amsmath}  % for math
\usepackage{amssymb}  % for math
\usepackage{hyperref} % for hyperlinks
\usepackage{cleveref}   % for referencing
\usepackage{color} % for color

\graphicspath{{./figures/}}

\begin{document}

\title{
Unveiling Exotic Magnetic Phases in Fibonacci Quasicrystalline Stacking of Ferromagnetic Layers through Machine Learning }
\author{Pablo S. Cornaglia }
\email[]{pablo.cornaglia@cab.cnea.gov.ar}
\affiliation{Centro Atómico Bariloche and Instituto Balseiro, CNEA, 8400 Bariloche, Argentina}
\affiliation{Consejo Nacional de Investigaciones Científicas y Técnicas (CONICET), Argentina}
\affiliation{Instituto de Nanociencia y Nanotecnología CNEA-CONICET, Argentina}

\author{Matias Nu\~nez}
 \affiliation{Consejo Nacional de Investigaciones Científicas y Técnicas (CONICET), Argentina}
	\affiliation{Instituto de Investigaciones en Biodiversidad y Medioambiente (INIBIOMA), Universidad Nacional del Comahue, Bariloche, Argentina}
 \affiliation{ Universidad de Ingenieria y Tecnologia- UTEC, Lima, Per\'u}

\author{D. J. Garcia}
\affiliation{Centro Atómico Bariloche and Instituto Balseiro, CNEA, 8400 Bariloche, Argentina}
\affiliation{Consejo Nacional de Investigaciones Científicas y Técnicas (CONICET), Argentina}
\date{\today}

\begin{abstract}

In this study, we conduct a comprehensive theoretical analysis of a Fibonacci quasicrystalline stacking of ferromagnetic layers, potentially realizable using van der Waals magnetic materials. We construct a model of this magnetic heterostructure, which includes up to second neighbor interlayer magnetic interactions, that displays a complex relationship between geometric frustration and magnetic order in this quasicrystalline system. To navigate the parameter space and identify distinct magnetic phases, we employ a machine learning approach, which proves to be a powerful tool in revealing the complex magnetic behavior of this system. We offer a thorough description of the magnetic phase diagram as a function of the model parameters. Notably, we discover among other collinear and non-collinear phases, a unique ferromagnetic alternating helical phase. In this non-collinear quasiperiodic ferromagnetic configuration the magnetization decreases logarithmically with the stack height. 

\end{abstract}

\maketitle

\section{Introduction}
The advent of two-dimensional (2D) materials has opened up a new chapter in the field of condensed matter physics, offering a rich platform for exploring novel phenomena \cite{ajayan2016two}. Among these, magnetic van der Waals (vdW) materials have attracted significant attention due to their unique magnetic properties and potential for integration into spintronic devices\cite{gong2017discovery,tian2019ferromagnetic,huang2017layer,bonilla2018strong,mak2019probing,doi:10.1021/acsnano.1c09150,burch2018magnetism}. These materials, characterized by their layered structure with weak interlayer bonding, offer the possibility of constructing heterostructures with tailored magnetic properties\cite{huang2020emergent}. For example, monolayer CrI$_3$  has been reported to be ferromagnetic\cite{huang2017layer} but the coupling between two layers can be ferromagnetic or antiferromagnetic depending on the type of stacking\cite{doi:10.1126/science.aav1937}.  
 The stacking of magnetic layers in a single heterostructure provides an opportunity to engineer the magnetic properties at the atomic scale, potentially leading to the realization of novel magnetic phases and spin textures.

Over the past few years, machine learning (ML) techniques have been increasingly utilized in condensed matter physics research due to their capabilities of dealing with large and complex data sets\cite{carrasquilla2017machine, broecker2017machine, nunez1, torlai2018neural,carleo2019machine}. Particularly, they provide a means to identify patterns and correlations within the data, which would be otherwise challenging or impossible to identify manually. This has allowed for new insights into several areas, such as phase transitions\cite{carrasquilla2017machine,wetzel2017unsupervised}, many-body localization\cite{schindler2017probing}, topological materials\cite{deng2017machine}, and visualization of band structure spaces from electronic structure databases\cite{nunez1}. In the realm of 2D materials\cite{nunez2}  and magnetic systems, ML techniques offer the possibility to help understand the magnetic properties of complex heterostructures.

In this theoretical study, we explore the magnetism in heterostructures having a quasicrystalline stacking of ferromagnetic layers. 
Quasicrystals are aperiodic structures that display sharp peaks in Bragg diffraction but lack translational symmetry \cite{shechtman1984metallic}. These materials have been extensively studied in the context of electronic transport, phonon propagation, and optical properties \cite{steinhardt1987physics,fujiwara2007quasicrystals,dubois2005useful}. 

The Fibonacci quasicrystals, often referred to as the ``fruit fly'' of quasicrystals, are a remarkable entity in the study of complex crystal structures \cite{jagannathan2020fibonacci}. This nickname draws a parallel between the role of fruit flies in genetic research and that of Fibonacci quasicrystals in the field of quasicrystal studies. Both are seen as relatively simple, model systems that help explore more complex phenomena.

The Fibonacci quasicrystals can be interpreted as the projection of a two-dimensional periodic crystal to a single dimension. This means that they can be formed by considering a subset of a regular, repeating structure in a space of higher dimensions, and then projecting this subset down to a lower-dimensional space. This process results in a pattern that lacks the perfect, repeating symmetry of a classic crystal, yet still exhibits a form of order, which is referred to as quasiperiodicity. They can be realized experimentally in a variety of systems, in particular heterostructures made from quasi-two-dimensional semiconducting layers \cite{rychly2015spin,rychly2016spin}, and potentially using vdW materials.

The study of magnetism in Fibonacci quasicrystals has been mainly focused on the nature of the magnetic excitations in a ferromagnetic ground state\cite{coelho2010quasiperiodic,costa2011partial,coelho2011transmission,costa2013band,machado2013static,rychly2015spin,rychly2016spin,grishin2013dissipative,lisiecki2019magnons,lisiecki2019reprogrammability}. Here we focus on the ground state properties of a model Fibonacci quasicrystalline stacking of ferromagnetic layers. The magnetic properties of this system are expected to be strongly influenced by the interlayer magnetic interactions. In particular, the frustration between first and second neighbor interlayer magnetic interactions is anticipated to play a key role in the emergence of non-collinear magnetic phases. In this study, we construct a model of this magnetic heterostructure, which includes up to second neighbor interlayer magnetic interactions. The resulting parameter space is explored using an unsupervised ML approach which combines Principal Component analysis (PCA) \cite{hotelling1933analysis,pearson1901liii,jolliffe2002principal,james2013introduction,abdi2010principal}, which is a dimensionality reduction technique, with Hierarchical Density-Based Spatial Clustering of Applications with Noise  (HDBSCAN),\cite{campello2013density} which is a clustering technique. We supplemented the powerful pattern recognition abilities of ML with a detailed analysis of the clusters obtained, constructing a simplified model that allows to describe the magnetic configuration on each phase. This approach allowed us to identify and characterize a rich phase diagram, which includes both collinear and non-collinear phases. 

The rest of the paper is organized as follows. In \cref{sec:model} we present the model and the methods used in this study. In \cref{sec:results} we present the results of the machine learning analysis and the phase diagram of the system. In \cref{sec:conclusion} we summarize our findings and discuss their implications.

\section{Model and Methods} \label{sec:model}
The construction of a Fibonacci stacking involves considering two types of interlayer bonds denoted as $S$ (short) and $L$ (long). This construction follows  an iterative procedure, also known as expansion rules:  

\begin{enumerate}
    \item Start the process with a bond $S$ connecting two layers.
    \item Apply an expansion step, where each $S$ is replaced with the bond $L$ and each $L$ is replaced by the sequence $LS$. 
    \item Repeat the expansion step described in the second rule for each new sequence of bonds until the targeted expansion step is reached.
\end{enumerate} 

By following these rules, a sequence of layers is generated, which displays quasicrystalline order. Each expansion step increases the number of layers and bonds by the number of $L$ bonds in the previous sequence.  At the $n$-th expansion step of a Fibonacci quasicrystal, the quantity of bonds is given by $F_n$ while the number of layers is $N=F_n+1$, where $F_n$ refers to the $n$-th Fibonacci number. The Fibonacci sequence of numbers is defined recursively through the formula $F_{n+2} = F_{n+1} + F_n$, starting from $F_0 = F_1 = 1$. 

Alternatively, the Fibonacci sequence of bonds can be constructed concatenating `words'. The starting words are $\omega_0=S$ and $\omega_1=L$. At step $m$ the word $\omega_m$ is given by the concatenation of the two previous words: $\omega_m=\omega_{m-1}\omega_{m-2}$ \cite{walter2009crystallography}. The first concatenation steps results in the words:

\begin{align*}
\omega_0&=S\\
\omega_1&=L\\
\omega_2&=\underbrace{L}_{\omega_1}\underbrace{S}_{\omega_0}\\
\omega_3&=\underbrace{LS}_{\omega_2}\underbrace{L}_{\omega_1}\\
\omega_4&=\underbrace{LSL}_{\omega_3}\underbrace{LS}_{\omega_2}\\
\omega_5&=\underbrace{LSLLS}_{\omega_4}\underbrace{LSL}_{\omega_3}\\
\omega_6&=\underbrace{LSLLSLSL}_{\omega_5}\underbrace{LSLLS}_{\omega_4}
\end{align*}

From this construction method it is easy to see that the number of times a given word $\omega_m$ appears in a subsequent word $\omega_n$ ($n\geq m$) is given by a Fibonacci number ($F_{n-m}$). For example, the arrangement $\omega_3=LSL$ appears $F_{3}=3$ times in the $\omega_6$ word.
\subsection{Fibonacci magnetic model}
To devise a magnetic model, we consider ferromagnetic layers connected by the bonds, and assume that the magnetic couplings between layers  are determined by the type of bonds separating the layers. It's important to note that two $S$ bonds are never adjacent to each other in the Fibonacci quasicrystal. 

Considering interactions at a separation of up to two layers, we identify four distinct types of couplings, namely: $\{S\text{ and }L\}$ for first neighbor layers and $\{LS (\equiv SL) \text{ and }LL\}$ for second neighbor layers. %The variety in these couplings can be attributed to differences in the interlayer atoms involved, as well as varying distances between the corresponding layers. 
We assume that the magnetic order within each layer is dominated by a ferromagnetic intralayer coupling. This allows us to treat each layer as a single classical magnetic moment, and the Fibonacci stacking as a one-dimensional chain where each site corresponds to a layer.
The energy that results from these considerations for a chain of length $N=F_n+1$ is given by 
\begin{align}\label{energy}
E &= \sum_{i=1}^{N-1} J(|R_{i+1}-R_i|)\vec{S}_i \cdot \vec{S}_{i+1}\nonumber \\
 &+ \sum_{i=1}^{N-2} J(|R_{i+2}-R_i|)\vec{S}_i \cdot \vec{S}_{i+2}   
\end{align}
where $\vec{S}_i$ is the magnetic moment of layer $i$, that we normalize to $1$, $R_i$ is the position of layer $i$, and we are using open boundary conditions. The coupling parameters are defined by: $J(S)=J_S$, $J(L)=J_L$, $J(S+L)=J(L+S)=J_{LS}$, and $J(L+L)=J_{LL}$. Since the minimum energy configuration is independent of the energy scale, we define normalized couplings $j_\alpha= J_\alpha/\sqrt{\sum_\alpha J_\alpha^2}$. This allows to focus the analysis on the relative strength of the couplings, rather than their absolute values, and it is used below for the graphical representation of the identified magnetic phases in parameter space. The actual values of the bond lengths are irrelevant for the determination of the magnetic phases. 

Although more complex models can be envisioned, this model already displays a rich phase diagram which stems from the quasicrystalline stacking and the frustration between first and second neighbor interlayer interactions. As we show below, the inclusion of an easy plane magnetic anisotropy does not change the results for the ground state magnetic configurations.

\subsection{Machine Learning Analysis} \label{sec:ml}
In order to determine the magnetic phases of the system we performed a random sampling of the coupling parameter space. For each set of coupling parameters, $J_\alpha$ ($\alpha=\{L, S, LS, LL\}$), we adjusted the orientation of the magnetic moments in order to minimize the energy of the configuration in finite-size Fibonacci stacks.

We used the Broyden-Fletcher-Goldfarb-Shanno algorithm (see \Cref{ap:energymin} for details) to obtain the configuration that minimizes the energy as defined by \cref{energy}. Due to the nature of the energy minimization procedure, in a handful cases the configuration obtained corresponded to a local minimum. However, the broad range of systems analyzed enabled us to differentiate the magnetic phases of the system.

To analyze the obtained magnetic configuration for each set of couplings, we calculated the Fourier transform of the magnetic moments along the stacking direction:
\[
    \vert S(q)\vert^2 = \left|\frac{1}{N}\sum_{j=1}^{N}e^{i \pi q j} \vec{S}_j \right|^2
\]
where $q = \frac{2 \ell}{N}-1$, with $\ell=0,\ldots N-1$. This quantity is independent of any global rotation of the magnetic moments as is the proposed model, and the normalized magnetization per layer can be written as $M=|S(q=0)|$. $|S(q)|^2$ was used to fuel a machine learning methodology aimed at identifying the magnetic phases of the system. 

We performed a dimensionality reduction using a Principal Component Analysis (PCA)  on the normalized\footnote{The data is normalized to have zero mean and a standard deviation of 1.} $|S(q)|^2$ data. PCA identifies axes in the high-dimensional data space that account for most of the data variation. This procedure can help to reduce the noise in the data and reduces the computational cost of the subsequent clustering scheme. \footnote{We also performed a dimensionality reduction using UMAP (Uniform Manifold Approximation and Projection) which is more robust to noise and outliers \cite{mcinnes2018umap}. However, the results of the clustering were similar to those obtained using PCA.}

After reducing the dimensionality, 
we utilized HDBSCAN \cite{campello2013density}, a density-based clustering algorithm, 
to identify distinct clusters within the data in this lower-dimensional space. HDBSCAN is renowned for its capability to detect clusters of varying densities. 
In the context of our work, data points assigned to a specific cluster are in proximity to each other which means that the magnetic configurations show a degree of similarity. Since this can imply that the data points belong to the same magnetic phase, it helped us guide the analysis of the phase diagram in parameter space.

Finally, we constructed a simplified model that allowed us to describe the magnetic configuration on each phase, and to reliably obtain the corresponding ground state configuration for each set of parameters.

\section{Results and Discussion} \label{sec:results}   
We navigated the parameter space, selecting a total of 60,000 sets of random values for the couplings $J_\alpha$ ($\alpha=\{L, S, LS, LL\}$). Each value of $J_\alpha$ was selected from a uniform distribution in the range $[-1, 1]$.  The minimum energy configuration for each set of parameters was obtained for systems defined by the $12$-th expansion stage of a Fibonacci quasicrystal (this corresponds to $N=F_{12}+1=145$ layers).
 The $|S(q)|^2$ was computed for all configurations and used as input for the machine learning analysis.

\subsection{Collinear Phases} \label{sec:col}
\begin{figure}
    \centering
    \includegraphics[width=0.475\textwidth]{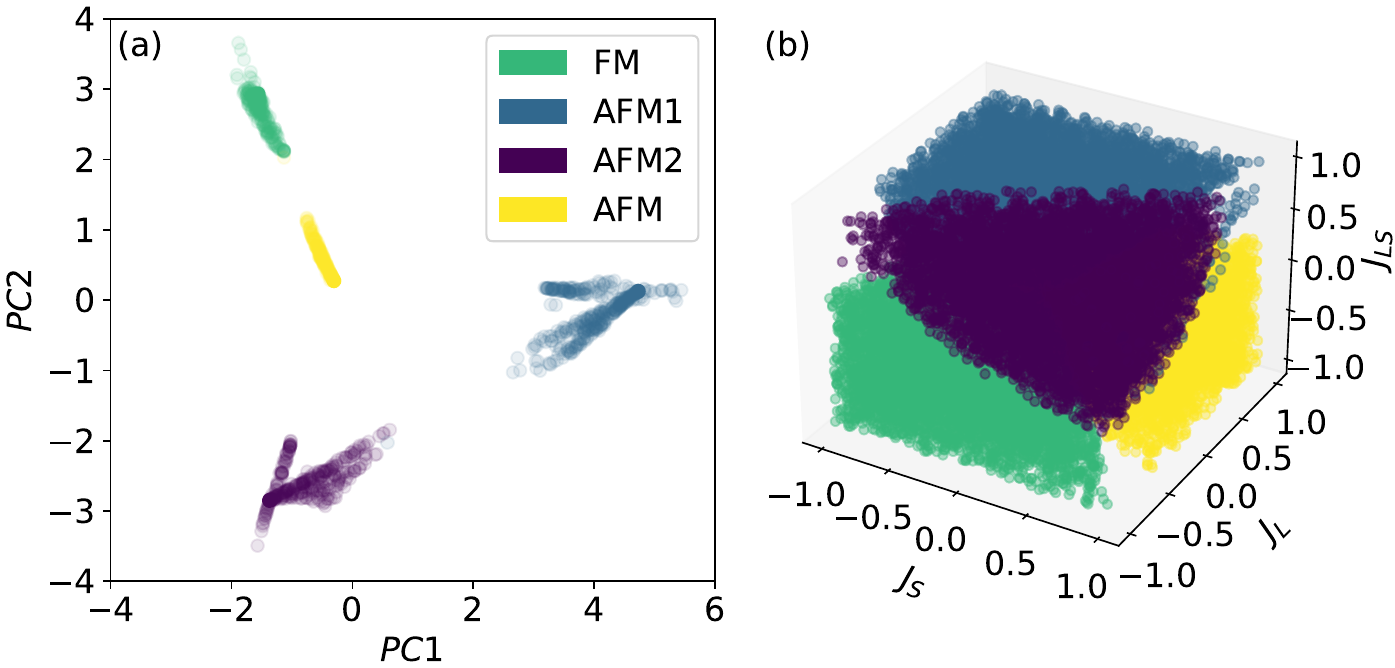}
    \caption{(a) Four clusters are obtained after a clustering of the full data set using HDBSCAN, keeping clusters with $1000$ data points or more.  The axes correspond to the two most relevant PCA axes. (b) The clusters  obtained correspond to collinear phases and are well separated in parameter space.} 
    \label{fig:clusters_col}
\end{figure}
We performed the PCA on the normalized $|S(q)|$ data for the 60000 points in the coupling parameter space. The dimensionality of the $|S(q)|^2$ data was reduced from $145$ to $65$, while retaining more than $95\%$ of the data variation. The first two principal components account for less than $10\%$ of the data variation, which is usually a signature of a complex data space.

Using a minimum threshold of $1000$ data points per cluster, HDBSCAN initially identified four clusters. \Cref{fig:clusters_col}(a) presents the points belonging to the obtained clusters in the space of the main two principal component axes. The clusters are well separated, which suggests that they correspond to qualitatively distinct groups of states or magnetic phases. The clusters are also separated in parameter space [see \ref{fig:clusters_col}(b)] and correspond indeed to distinct collinear phases, as it can be checked by inspection of the magnetic moment configurations. The four collinear phases are: ferromagnet (FM), staggered antiferromagnet (AFM), antiferromagnet with the magnetic moments across $S$ bonds ferromagnetically aligned (AFM1), and antiferromagnet with the magnetic moments across the $L$ bonds ferromagnetically aligned (AFM2).

In order to analyze the identified phases, we first point out that the energy of the system is symmetric when the magnetic moments at odd (or equivalently even) layer number are inverted and the sign of the of nearest neighbor couplings $J_S$ and $J_L$ is changed [$\vec{S}_i \to (-1)^{i}\vec{S}_i$, $J_S\to-J_S$, and $J_L\to -J_L$]. This reversal of the magnetic moments doesn't modify the interaction energy between second neighbor magnetic moments, but it does change the interaction sign for the first neighbor ones. This symmetry can be seen in \cref{fig:clusters_col}(b) where the region in parameter space associated with the antiferromagnetic cluster AFM can be obtained from the ferromagnetic one by applying the transformation: $J_S\to-J_S$, and $J_L\to -J_L$. The same also holds for the phases AFM1 and AFM2. The inversion of the odd magnetic moments results in a shift by $1$ for the value of $q$ in $|S(q)|^2$. Although the PCA transforms the $|S(q)|^2$ data, hints of this symmetry can be observed for the clusters in principal component space [\cref{fig:clusters_col}(a)]. 
\subsubsection*{Magnetic structure of the collinear phases}
\begin{figure}
    \centering
    \includegraphics[width=0.475\textwidth]{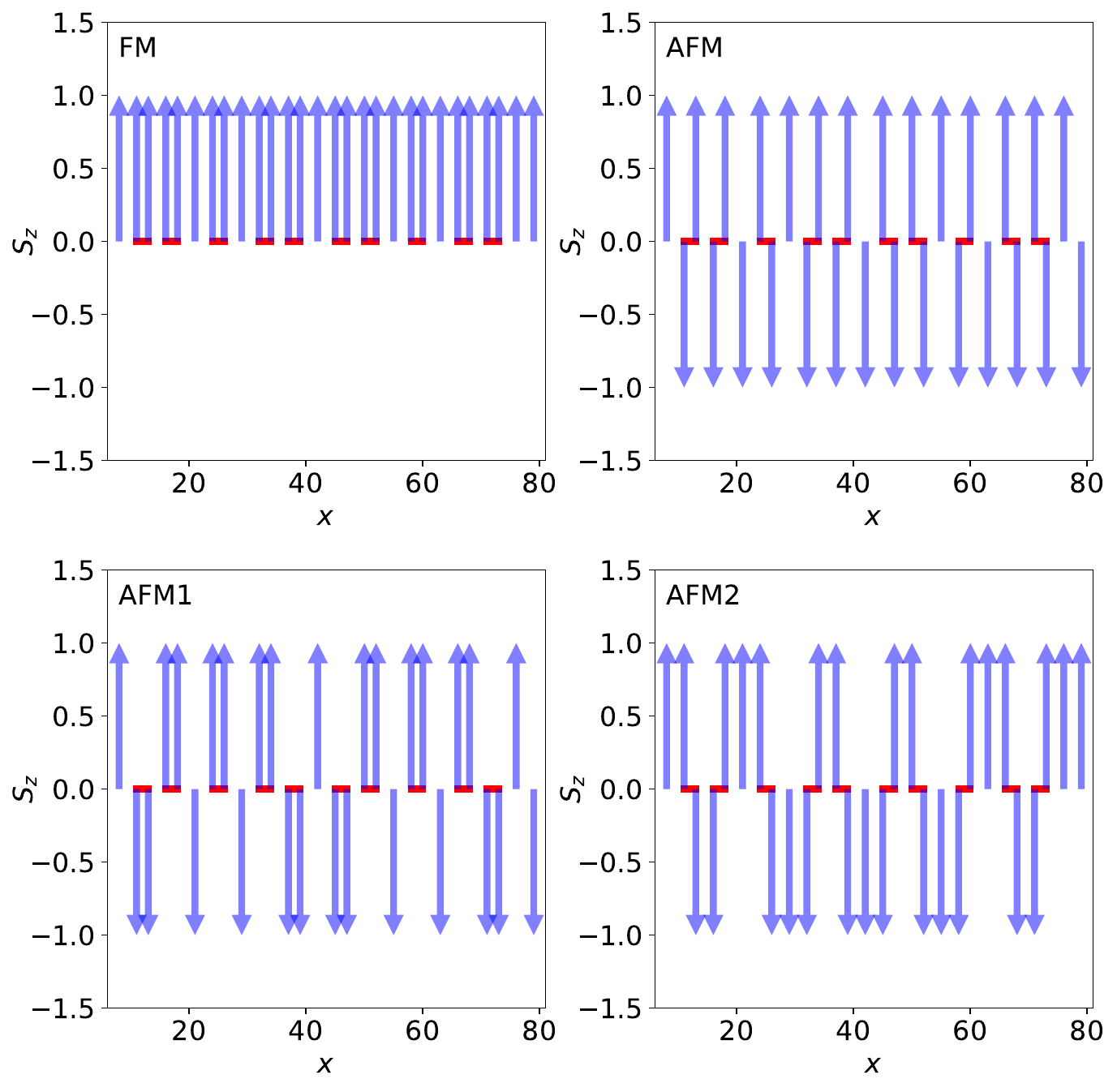}
    \caption{Schematic representation of the different collinear ground states. The magnetic moment on each site (layer) is represented by a blue arrow.  The stacking direction is $x$, and the distances for $S$ and $L$ bonds were set, for plotting purposes only, to $2$ and $3$, respectively. The segments connecting arrows along the $x$-axis correspond to $S$ bonds. }
    \label{fig:states_col}
\end{figure}
In the FM phase all magnetic moments are aligned, and the normalized magnetization is $M=1$.
In the AFM phase the magnetic moments are staggered along the stacking direction, and the magnetization is $M=0$.
The AFM1 phase is collinear and has zero magnetization as the AFM phase, but there is a ferromagnetic alignment of the magnetic moments across the $S$ bonds while the alignment of the magnetic moments across $L$ bonds is antiferromagnetic.  This magnetic structure can be obtained treating the two ferromagnetic layers across an S bond as a unit and building a staggered antiferromagnet. For example in this Fibonacci sequence ($\omega_6$):
\begin{align*}
    \bullet  L \underbrace{\bullet S\bullet}_\square L\bullet L\underbrace{\bullet S\bullet}_\square L\underbrace{\bullet S\bullet}_\square L\bullet L\underbrace{\bullet S\bullet }_\square L\bullet L\underbrace{\bullet S\bullet}_\square
\end{align*}
the dots correspond to layers.  The pairs of sites across an S bond are replaced by an effective site (square shape). 
This leads to 
\begin{align*}
    \bullet  L \square L\bullet L\square L\square L\bullet L\square L\bullet L\square
\end{align*}
and setting a staggered magnetization for dot and square sites leads to the AFM1 state:
\begin{align*}
    \uparrow  \boxed{\downarrow\downarrow} \uparrow \boxed{\downarrow\downarrow}\boxed{\uparrow\uparrow} \downarrow \boxed{\uparrow\uparrow} \downarrow \boxed{\uparrow\uparrow}
\end{align*}
where the boxed arrows correspond to magnetic moments separated by an $S$ bond.
By construction the sequence of squares and dots has a Fibonacci quasicrystalline structure, and consequently, this is also the case for the magnetic state.

An analogous construction can be made for the AFM2 states. The main difference is that for this phase the magnetic moments on sites across an $L$ bond are ferromagnetically aligned. Therefore, the sites along the chain can be grouped in sets of two sites (across an L bond) and three sites (across two consecutive L bonds) that have the same magnetic moment orientation. Alternatively, an AFM2 state can be obtained from the AFM1 state by flipping the odd site magnetic moments using the above-mentioned symmetry of the model.

The four collinear phases are schematically represented in \cref{fig:states_col}, where the stacking direction is along the $x$ axis, the magnetic moments are represented by arrows, and we have used the rotational symmetry of the model to align the magnetic moments along the $z$ axis.

\subsubsection*{Energy of the collinear phases}
The energy of the collinear ferromagnetic phase and the three collinear antiferromagnetic phases can be written using the couplings $J_\alpha$ and the count of each type of coupling. In a Fibonacci quasicrystal at the $n$-th expansion step (for $n\geq 4$), the count is as follows: there are $F_{n-1}$ bonds of type $L$, $F_{n-2}$ bonds of type $S$, $F_{n-2}$ and $F_{n-3}\pm1$ bonds of type $LS$ and $LL$, respectively.  Thus, the total energy of the phase $\beta$ can be computed as:
%: $F_{n-1}$ $L$ bonds, $F_{n-2}$ $S$, $2F_{n-2}$ $LS$ bonds, and $F_{n-3}$ $LL$ bonds.

\begin{align*}
E_\beta(n) &\simeq \lambda_\beta^L J_L F_{n-1} + \lambda_\beta^S J_S F_{n-2}\\ &+ \lambda_\beta^{LS} J_{LS} 2F_{n-2}+ \lambda_\beta^{LL}J_{LL} F_{n-3}, 
\end{align*}
where the coefficients $\lambda_\beta^\alpha$ are defined as follows:

\begin{tabular}{c|c|c|c|c}
    $\lambda_\beta^\alpha$       & $\alpha = S$ & $\alpha = L$ & $\alpha = LS$ & $\alpha = LL$ \\
\hline
$\beta = \text{FM}$   & + & + &  + &  + \\
\hline
$\beta = \text{AFM}$  & - & - &  + &  + \\
\hline
$\beta = \text{AFM1}$ & + & - &  - &  + \\
\hline
$\beta = \text{AFM2}$ & - & + &  - &  + \\
\end{tabular}

In the thermodynamic limit ($n\to \infty$), the energy per site (layer) $E_\beta=\lim_{n\to \infty}E_\beta(n)/F_n$ can be denoted in terms of the golden ratio $\phi=\lim_{n\to \infty} F_n/F_{n-1}$:
\begin{align} \label{eq:colen}
E_\beta =&\lambda_\beta^L J_L\phi^{-1}  + \lambda_\beta^S J_S \phi^{-2}\nonumber \\ &+ \lambda_\beta^{LS} J_{LS} 2\phi^{-2}+ \lambda_\beta^{LL}J_{LL} \phi^{-3}.
\end{align}

We verified that the collinear magnetic structures obtained are robust and that their energy converges to the value given by \cref{eq:colen} as the system size is increased for a given set of parameters. Additionally, the magnetization of the antiferromagnetic states converges to zero.

\subsection{Non-Collinear Phases}
Of the 60000 sets of coupling parameters, $32319$ were assigned to one of the four collinear phases, while the rest could not be assigned to a cluster using the threshold of 1000 data points. Reducing the cluster size threshold resulted in an increasing number of additional clusters that could not be easily distinguished from each other or classified as magnetic phases. 

\begin{figure}
    \centering
    \includegraphics[width=0.475\textwidth]{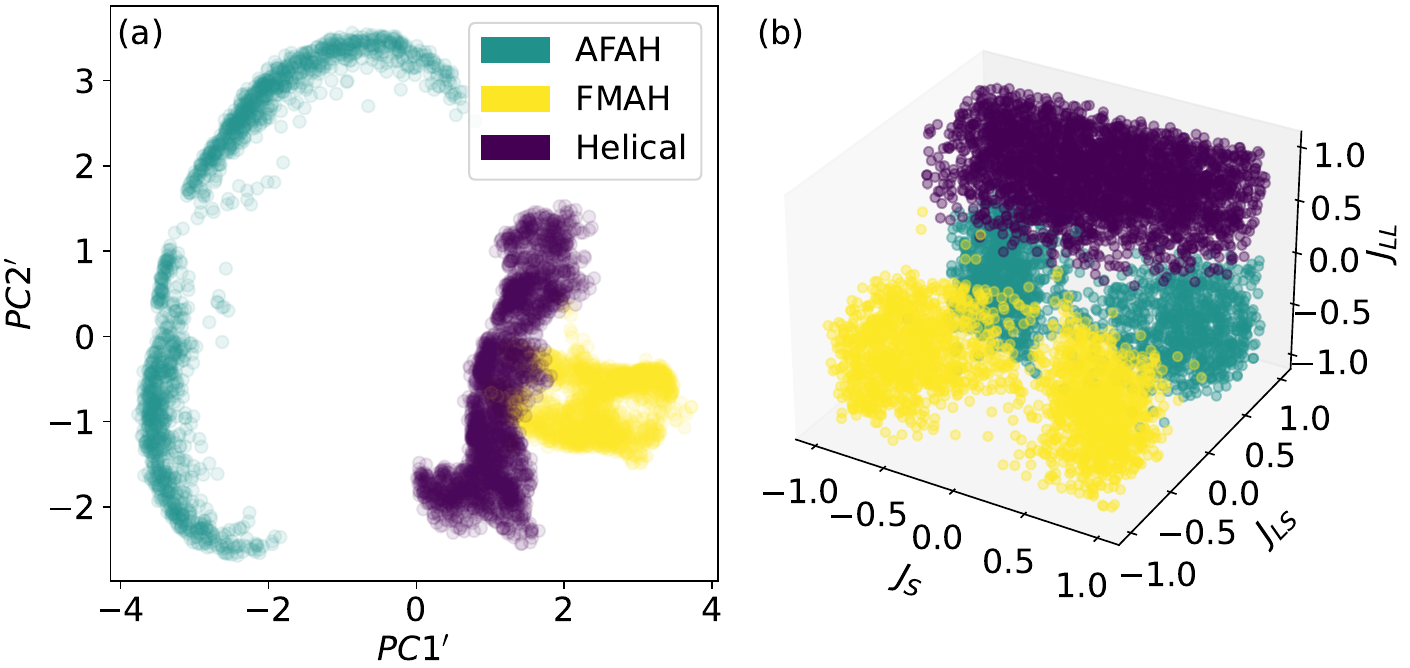}
    \caption{(a) Clustering using a reduced dataset without the points corresponding to collinear phases. Keeping clusters with at least $500$ points, three clusters are identified, that correspond to non-collinear phases. The principal component axes differ from those of \cref{fig:clusters_col}. (b) The clusters correspond to separated regions in parameter space. }
    \label{fig:clusters_nocol}
\end{figure}

To simplify the identification of additional phases, we conducted a second stage of analysis, excluding data points associated with the previously identified collinear phases. Performing the PCA and HDBSCAN procedure on the reduced dataset, using minimum cluster size of 500 data points, led to the identification of three distinct 
clusters. The clusters are well separated in the principal component space [see \Cref{fig:clusters_nocol}(a)] and correspond to distinct
non-collinear phases [see \Cref{fig:clusters_nocol}(b)]. 
 The three non-collinear phases are: a ferromagnetic phase with an alternating helical structure (FMAH), a helical phase (Helical), and an antiferromagnetic phase with an alternating helical structure (AFAH). The criterion to name the phases is based on the magnetic structure of each phase, which is schematically represented in \cref{fig:states_nocol} and will be discussed below. 

\subsubsection*{Magnetic structure of the non-collinear phases}
\begin{figure}
    \centering
    \includegraphics[width=0.475\textwidth]{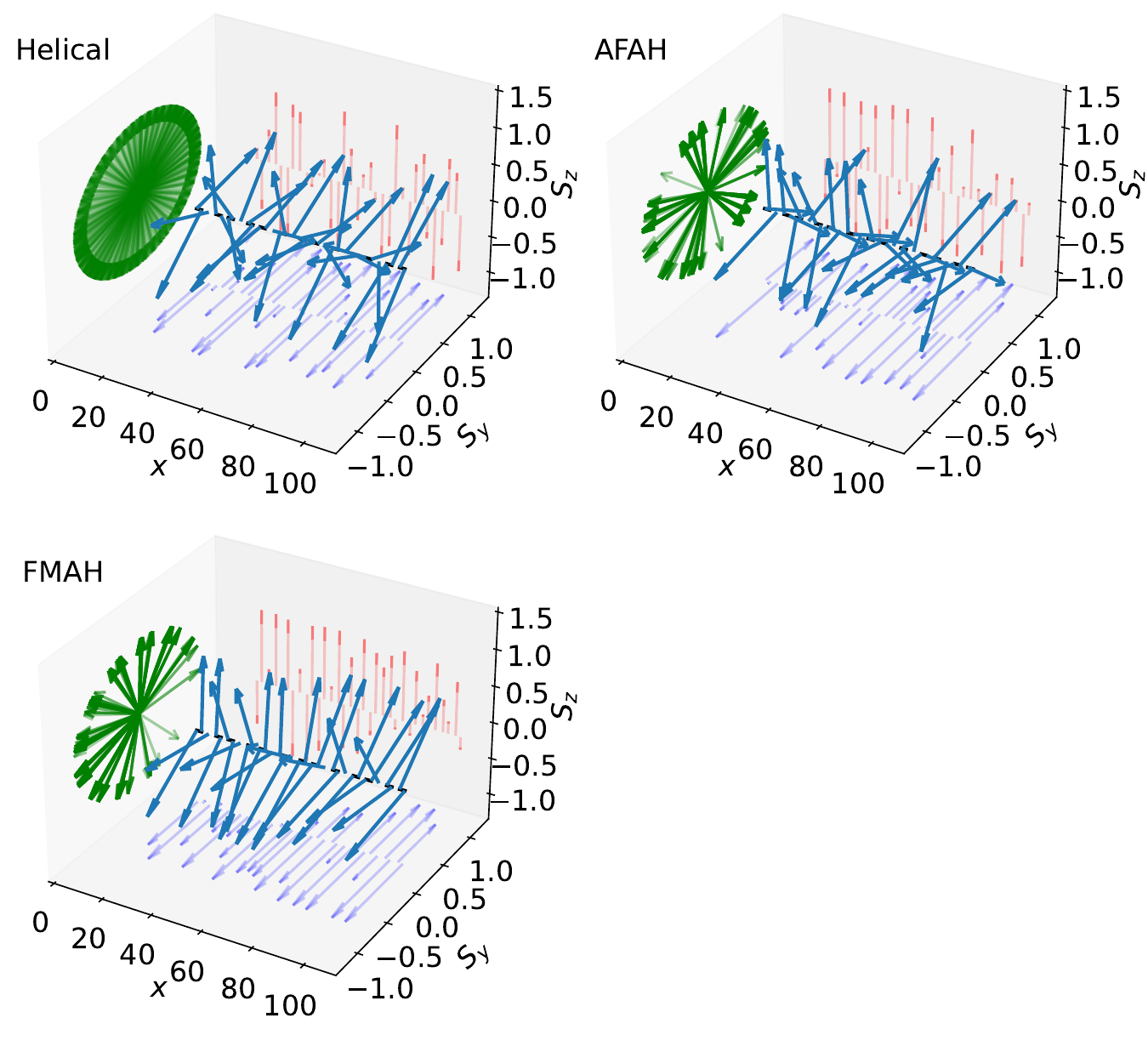}
    \caption{Schematic representation of the different non-collinear ground states. The magnetic moment on each site (layer) is represented by a blue arrow.  The stacking direction is $x$, and the distances for $S$ and $L$ bonds are set, for plotting purposes only, to $2$ and $3$, respectively. The projection of the magnetic moments on the $S_y$-$S_z$, $x$-$S_y$, and $x$-$S_z$ planes are represented using green, light blue, and red arrows, respectively. The segments connecting arrows along the $x$-axis correspond to $S$ bonds. The configurations depicted correspond to a subset of the sites in systems of 988 sites, for which the energy minimization was performed.}
    \label{fig:states_nocol}
\end{figure}

Representative states of three non-collinear phases are schematically represented in \cref{fig:states_nocol}, where the stacking direction is along the $x$ axis, the magnetic moments are represented by blue arrows, and we have used the rotational symmetry of the model to align the magnetic moments perpendicular to the $x$ axis.  

A closer examination of these states unveils a fundamental difference between the Helical and alternating helical phases. Specifically, in the alternating helical phases, the magnetic moments across a double $L$ bond align in parallel. This characteristic is not present in the Helical phase. Another feature that distinctly separates the phases is the chirality of the states:
\begin{equation}
    \vec{\kappa} = \frac{1}{N-1}\sum_{i=1}^{N-1} \vec{S}_i\times \vec{S}_{i+1}.
\end{equation}

The Helical phase is characterized by a finite chirality and vanishing magnetization. The FMAH phase is characterized by a finite magnetization ($0<M<1$) and vanishing chirality. The AFAH phase is characterized by vanishing magnetization and vanishing chirality.

\subsubsection*{Energy of the non-collinear phases}
In layered systems, a non-collinear (helical) phase typically emerges due to a frustration between first and second neighbor interlayer couplings. This requires the second neighbor couplings to be antiferromagnetic (greater than zero). In the Fibonacci quasicrystal, there are two first-neighbor couplings and two second-neighbor couplings, which could lead to anticipate two characteristic angles for the helix, one for each type of bond. Under this assumption, the energy of the helical phases would be given by an extension of the expression for the energy of the collinear phases \cref{eq:colen}:

\begin{align*}
E^{2\theta} &= J_L \phi^{-1}\cos(\theta_S)  + J_S \phi^{-2}\cos(\theta_L)\\ &+ J_{LS} 2\phi^{-2}\cos(\theta_S+\theta_L) + J_{LL} \phi^{-3} \cos(2 \theta_L),
\end{align*}

Here, $\theta_\ell=\arccos(\vec{S}_i\cdot \vec{S}_{i+1})$ represents the angle between magnetic moments in nearest neighbor layers across an $\ell=\{L,S\}$ bond.
 We have assumed coplanar magnetic moments, where the relative angle between second neighbors is determined by the sum of the relative angles in between. 

 This two-angle model captures the collinear phases exactly since $\theta_S=0$ or $\pi$ and $\theta_L=0$ or $\pi$ for a collinear phase. Although it also provides a first approximation to the energy of the system, it falls short in depicting the intricacies of the non-collinear phases in the Fibonacci quasicrystal.

Generally, the relative angle for two nearest-neighbor magnetic moments is expected to be influenced not just by the bond type, but by the surrounding environment of the bond as well. Different angles are generally expected for an $L$ bond flanked by two $S$ bonds ($S\mathbf{L}S$ configuration), and for a central $L$ bond in the $L\mathbf{L}S$ or $S\mathbf{L}L$ configurations. A similar situation is anticipated for the $S$ bonds, where the angle for the $LL\mathbf{S}LL$ configuration and for the central bond in the $LL\mathbf{S}LS$ and $SL\mathbf{S}LL$ configurations may differ. To accurately describe the ground state, at least two relative angles for each type of bond ($\theta_i^\ell$, for $i=1,2$ and $\ell = S,L$) are generally needed.

For a Fibonacci quasicrystal constructed after $n$ expansion steps, we find $F_{n-5}\pm1$ $LL\mathbf{S}LL$ configurations, $F_{n-4}\pm 1$ $LL\mathbf{S}LS$, $SL\mathbf{S}LL$, and $L\mathbf{L}S$ configurations, and $F_{n-3}\pm 1$ $L\mathbf{L}S$ and $S\mathbf{L}L$ configurations. 
Using this four-angles model, the energy of the non-collinear phases is given in the thermodynamic limit by:

\begin{align} \label{eq:4angles}
E_{\pm}^{4\theta} =&J_S \left( \phi^{-5} \cos(\theta^{S}_{1}) +2 \phi^{-4}  \cos(\theta^{S}_{2})\right)\nonumber\\
 &+ J_L \left(\phi^{-4} \cos(\theta^{L}_{1}) + 2 \phi^{-3}  \cos(\theta^{L}_{2})\right)\nonumber\\
 &+ J_{LS}  (2 \phi^{-5} \cos(\theta^{S}_{1} + \theta^{L}_{2}) + 2\phi^{-4} \cos(\theta^{L}_{1} + \theta^{S}_{2})\nonumber\\ 
&+2\phi^{-4} \cos(\theta^{S}_{2} + \theta^{L}_{2}))\nonumber\\
    &+ J_{LL} \phi^{-3}  \cos(\theta^{L}_{2} \pm \theta^{L}_{2}),
\end{align}
where, in the last term, we have allowed the possibility that the angle across a double $L$ bond may be twice the angle across a single $L$ bond or zero.
It's crucial to account for both positive and negative signs for the rotation angle
 after two consecutive $L$ couplings in order to capture all possible ground state configurations. A negative sign is associated with an opposing rotation angle of the magnetic moments in the $L$ bonds of an $LL$ double bond, while a positive sign indicates rotation in the same direction. Consequently, a negative sign implies a change in rotation direction, from clockwise to counterclockwise or vice versa, while a positive sign indicates continued rotation in the same direction.

We find that the Helical phase corresponds to states described by the $+$ sign while the AFAH and the FMAH correspond to the $-$ sign. 

The four-angle description allows a precise calculation of the ground state configuration energy, for any set of couplings, minimizing the energy $E_\pm^{4\theta}$ with respect to the four angles $\theta^{\ell=S,L}_{i=1,2}$. The minimization was performed using the {\it Differential Evolution} method \cite{storn1997differential} which is a fast stochastic method to obtain global minima. The two possible signs in the energy $E_{\pm}^{4\theta}$ of \cref{eq:4angles} were considered, and the minimum energy solution retained after comparing the two cases ($+$ and $-$). This approach turned out to be much faster than the energy minimization of finite size systems using the BFGS algorithm, and it was more reliable for the identification of the ground state configuration. Once the angles $\theta^{\ell=S,L}_{i=1,2}$ that minimize the energy are obtained, the magnetic configuration can be constructed for systems of arbitrary size and the magnetization and chirality can be readily calculated. Additionally, the four-angles solution is an excellent starting point to perform a full energy minimization of the system using \cref{energy}. For a given set of couplings, the ground state energy obtained from $E_\pm^{4\theta}$ generally remains within approximately 0.1\% of the ground state energy obtained using \cref{energy} for the largest systems considered ($N=4182$).

A substantial number of data points were not incorporated by HDBSCAN into the three clusters, which underscores the inherent complexity of identifying non-collinear phase structures. Nonetheless, once the seven distinct phases were identified, the four-angles model allowed the description of the magnetic configurations for all data points. 

While the four-angles model can be refined considering longer range environments for each bond, which leads to consider more possible angles for each bond, it is enough to characterize the phases of the system. 
\subsubsection*{Finite size effects}
We analyzed possible finite size effects on the non-collinear phases minimizing the energy for systems of up to 4182 sites. The four-angles model provides an excellent approximation for the ground state energy and magnetic configuration. However, as the system is increased, an increasing number of angles is in general needed to provide a full description of the magnetic configuration. 

The chirality of the alternating helical phase states converges to zero as the system size is increased, while it converges to a finite value for the Helical phase. The magnetization of the AFAH and Helical phases converges to zero, but magnetization in the FMAH phase shows a peculiar behavior as the system size is increased. 

\begin{figure}
    \centering
    \includegraphics[width=0.45\textwidth]{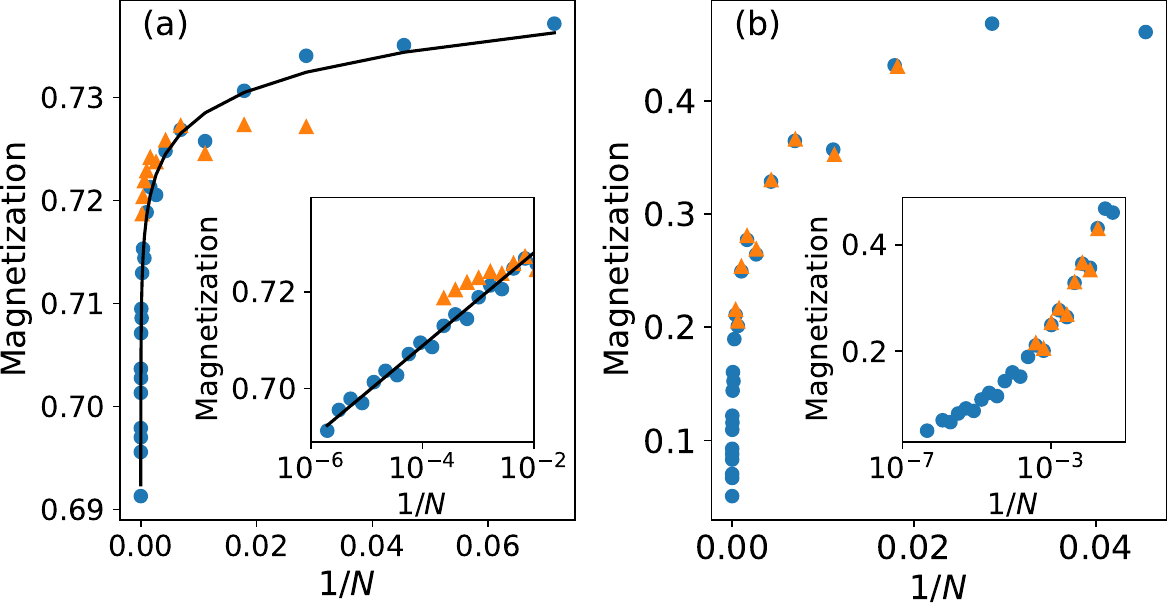}
    \caption{Scaling of the magnetization with the system size in the FMAH phase. The magnetization is calculated using the 4 angles model (disks) and the full minimization (triangles). (a) The couplings are $J_S=-0.26$, $J_{LS}=0.16$, and $J_{LS}=  -0.66$ $J_{LL}=-0.68$. (b) The couplings are $J_S=0.66$, $J_{LS}=-0.57$, and $J_{LS}=  -0.32$ $J_{LL}=-0.36$.}
    \label{fig:scalingmag}  
 \end{figure}

The results for the magnetization of a state in the FMAH phase, as a function of the inverse system size, calculated using both the four-angles model and a full minimization of the energy are presented in \cref{fig:scalingmag}. Interestingly, a very slow (logarithmic) decay of the magnetization is observed, which seems to indicate a vanishing magnetization for large enough systems. However, the decay of the magnetization with system size is so slow that exponentially large systems would be needed to observe a vanishing of the magnetization, assuming the trend continues as the system size is increased. For the set of parameters in our dataset where the systems showed the fastest decay of the magnetization, a change in the behavior is observed showing a decrease in the decay speed for large enough systems [see \cref{fig:scalingmag} (b)]. 

These results suggest that the magnetization of any experimentally achievable system with parameters in the FMAH phase will be finite.

\subsection{Phase Diagram}
\begin{figure*}
    \centering
    \includegraphics[width=\textwidth]{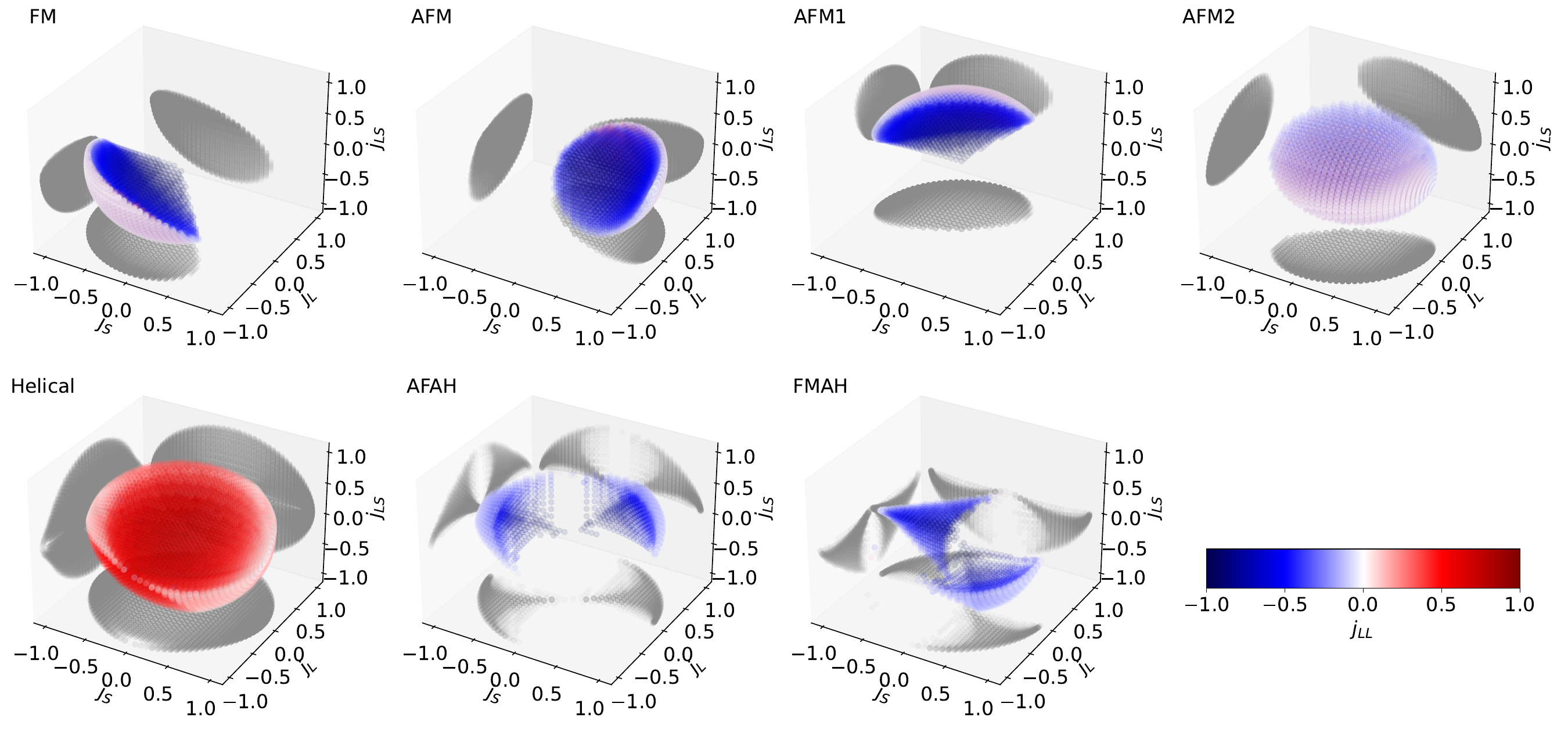}
    \caption{Magnetic phases of the magnetic Fibonacci quasicrystal. The coupling constants are normalized to the sum of their squares. Negative (positive) values are associated with (anti-)ferromagnetic  couplings. The color code represents the value of the $j_{LL}$ coupling. The phases indicated in the figure are FM: ferromagnetic, AFM: antiferromagnetic, AFM1: antiferromagnetic with the magnetic moments across $S$ bonds ferromagnetic, AFM2: antiferromagnetic with the magnetic moments across the $L$ bonds ferromagnetic, Helical: helical structure, AFAH: antiferromagnetic with an alternating helical structure. FMAH: ferromagnetic with an alternating helical structure. }
    \label{fig:phasediagram}
\end{figure*}

In order to plot the different phases of the system, we selected a uniform grid of points in parameter space and obtained the ground state configuration using the four-angles model. 
The points in parameter space were classified according to the value of the four angles that minimized the energy, the chirality and the magnetization. These features are presented in \Cref{tab:phases} for the seven magnetic phases identified. 
\begin{table}[t]
  \caption{Main characteristics of the magnetic Fibonacci quasicrystal phases.}
  \begin{tabular}{ccccccc}
      \hline
      Phase & Collinear? & $\theta^L_1,\theta^L_2$ & $\theta^S_1,\theta^S_2$ &  $M$ & $\kappa$ \\
      \hline
      FM    &     Yes    &            0 &            0 &      $1$ &      0   \\
      AFM   &     Yes    &            $\pi$ &            $\pi$ &      0 &        0   \\
      AFM1  &     Yes    &            $\pi$ &            0 &      0 &        0  \\
      AFM2  &     Yes    &            0 &            $\pi $&      0 &        0\\ 
\hline
      FMAH    &     No     &             &           &   $\lesssim 1$ & 0\\
      AFAH    &     No     &             &           &       0 &       0  \\ 
      Helical  &  No    &             &           &       0 &$\sim 1 $ \\ 
\hline
  \end{tabular}\label{tab:phases}
\end{table}

Only the Helical phase presents a non-zero chirality, while the FM and FMAH phases are the only ones having a finite magnetization in the ground state configuration.
In the FM phase the four angles of the model are zero ($\theta^L_1=\theta^L_2=\theta^S_1=\theta^S_2=0$) which corresponds to a saturated magnetization, while in the AFM phase, the four angles are $\pi$ ($\theta^L_1=\theta^L_2=\theta^S_1=\theta^S_2=\pi$). 
In the AFM1 phase, the angles are $\theta^L_1=\theta^L_2=\pi$ and $\theta^S_1=\theta^S_2=0$, which corresponds to the behavior described in \cref{sec:col}, where the magnetic moments across an $S$ bond are parallel, while the ones across an $L$ bond are antiparallel.
In the AFM2 phase, the situation is reversed compared to the AFM1 phase as the angles are $\theta^L_1=\theta^L_2=0$ and $\theta^S_1=\theta^S_2=\pi$.
Finally, in the non-collinear phases the angles are generic, and the phases can be distinguished by a finite unsaturated magnetization and zero chirality in the FMAH phase, a finite chirality and zero magnetization in the Helical phase, and a zero magnetization and zero chirality in the AFAH phase. 

All the magnetic phases identified are presented in~\cref{fig:phasediagram} showing on each subplot points in the normalized space of couplings $(j_S, j_L,j_{LS})$ that correspond to a given phase.  The color code represents the value of the $j_{LL}$ coupling. 
The FM phase, with all magnetic moments aligned, includes the systems with all couplings ferromagnetic ($j_\alpha< 0$). This is a collinear phase with saturated magnetization ($M=1$) and zero chirality ($\kappa=0$). The AFM phase features a staggered magnetization along the stacking direction and contains systems with antiferromagnetic first-neighbor couplings ($j_{L}, j_{S}> 0$) and ferromagnetic second neighbor couplings ($j_{LS}, j_{LL}<0$). Both the magnetization and the chirality vanish in this phase.

The FM and AFM phases in this system do not differ from the corresponding phases in a periodic stacking of ferromagnetic layers. The collinear AFM1 and AFM2 phases, however, are unique to the Fibonacci quasicrystal. Systems with ferromagnetic $j_S$ and $j_{LL}$ and antiferromagnetic $j_L$ and $j_{LS}$ couplings are found in the collinear AFM1 phase, while the AFM2 phase is mainly composed by systems with ferromagnetic $j_L$ and $j_{LL}$ and antiferromagnetic $j_S$ and $j_{LS}$ couplings. 

The Helical phase is in an antiferromagnetic $J_{LL}$ coupling region, while in the FMAH and AFAH phases, $J_{LL}$ is negative (ferromagnetic). The sign of $J_{LS}$ is different in the AFAH ($J_{LS}>0$) and FMAH ($J_{LS}<0$) phases.
The two lobes in the FMAH phase are related by the odd site inversion symmetry discussed above.  One of its lobes in parameter space is continuously connected to the ferromagnetic phase, while the other is connected to the AFM phase.
\subsubsection*{Phase classification using a neural network}
The complexity of the phase diagram significantly complicates a priori determination of the phase to which a specific parameter set belongs. However, by employing a trained neural network that uses classified points and normalized parameters as input, we successfully developed a classifier capable of predicting the phase with an accuracy exceeding $97.5\%$. Specifically, we trained a Multi-Layer Perceptron (MLP) model using the MLPClassifier function from the scikit-learn library in Python \cite{pedregosa2011scikit}\footnote{The model configuration comprised three hidden layers of sizes 10, 4, and 10, respectively. The 'tanh' activation function and the 'lbfgs' solver for weight optimization were employed. The model underwent a maximum of 5000 iterations with an L2 penalty parameter (alpha) of 0.0001 utilized for regularization.}. We trained the MLP model on 80\% of the data points, leaving the remaining 20\% for testing. The code to use the neural network is available as a Jupyter notebook\footnote{The Jupyter notebook to retrieve the Neural Network data and use it to predict the magnetic phase is available at: \url{https://doi.org/10.5281/zenodo.8185022}}. 
\begin{table}[h]
    \centering
    \caption{Confusion Matrix}
    \begin{tabular}{cccccccc}
    \hline
     & Helical & AFM & AFM2 & FMAH & AFAH & AFM1 & FM \\ 
    \hline
    Helical &  97.0 & 0.3 & 0.5 & 0.6 & 0.6 & 0.4 & 0.6 \\
    AFM & 0.6 & 98.9 & 0.0 & 0.2 & 0.2 & 0.0 & 0.1 \\  
    AFM2 &  1.3 & 0.0 & 97.9 & 0.2 & 0.6 & 0.0 & 0.0 \\ 
    FMAH & 1.5 & 1.5 & 0.9 & 94.5 & 0.0 & 0.9 & 0.8 \\  
    AFAH &  1.4 & 1.0 & 3.0 & 0.0 & 92.3 & 2.1 & 0.3 \\ 
    AFM1 &  1.1 & 0.0 & 0.0 & 0.0 & 0.7 & 98.2 & 0.0 \\ 
    FM & 1.0 & 0.0 & 0.0 & 0.5 & 0.1 & 0.0 & 98.3 \\  
    \hline
    \end{tabular}\label{tab:confusion}
\end{table}

In a more comprehensive assessment of our MLP model, we utilized a confusion matrix \Cref{tab:confusion}.
The diagonal elements correspond to the percentage of correctly classified data points (predicted label equals the true label), and the off-diagonal elements represent the percentage of misclassified data points. Each row signifies the true label, while each column indicates the predicted label. The MLP model displays good performance across all categories, and can be used for a quick exploratory analysis of the phase diagram. It is, however, important to bear in mind that the model's accuracy decreases near the phase boundaries. 
The Multilayer Perceptron (MLP) model, although powerful, does not fully capture the fine details in situations where the phase transition is a continuous process and the differences between phases are subtle. In these cases, the four-angles model is a more reliable tool to identify the phase of a given set of couplings.

\subsubsection*{Phase boundaries}
\begin{figure}
    \centering
    \includegraphics[width=0.475\textwidth]{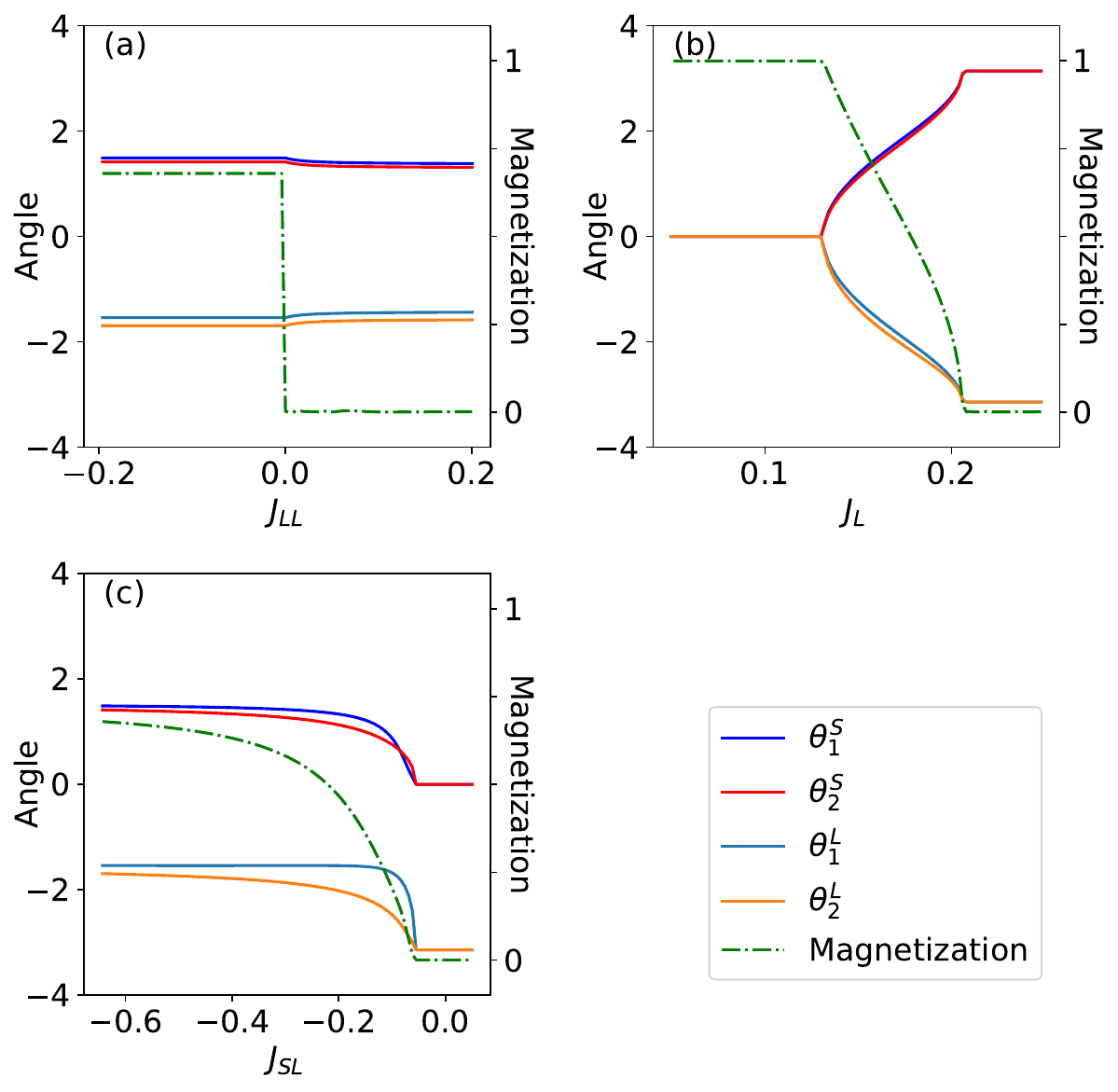}
    \caption{ Magnetization and angles of the 4 angle model, (a) for $J_S=-0.26$, $J_{L}=  0.16$, and $J_{LL}= -0.68$, as a function of $J_{LL}$. The system transition from the FMAH phase to the Helical phase at $J_{LL}=0$, where the magnetization vanishes discontinuously, while the angles show a continuous behavior.  (b) Same as (a) for $J_S=-0.26$, $J_{LS}=  -0.65$, and $J_{LL}= -0.68$ as a function of $J_L$. The system transitions from the ferromagnetic to the antiferromagnetic phase through the FMAH ferromagnetic non-collinear phase. 
     (c) Same as (a) for $J_S=-0.26$, $J_{LS}=0.16$, and $J_{LS}=  -0.65$ as a function of $J_{LS}$.  There is a continuous transition from the FMAH phase to the AFM2 phase.} 
    \label{fig:paths}
\end{figure}

To analyze the behavior of the system close to the phase boundaries, we used the four-angles model, which allows us to calculate reliably the ground state configuration for any set of couplings. \Cref{fig:paths} presents the magnetization and the angles of the four-angles model along three different paths in the parameter space. The transitions as a function of the parameters are in general continuous. The only exception is the transition from the FMAH phase to the Helical phase, which is discontinuous. The magnetization vanishes discontinuously, while the angles show a continuous behavior. 

\section{Conclusion} \label{sec:conclusion}
In this paper, we presented a study of the magnetic properties of a Fibonacci quasicrystalline magnetic heterostructure. We utilized a reduced Heisenberg model incorporating first- and second-neighbor couplings along the stacking direction to describe the system's magnetic frustration effects.

An unsupervised machine learning methodology allowed for the exploration of the parameter space, aiding in the identification of the system's ground state magnetic configurations. The use of a simplified model, defined by four distinct angles representing the nearest neighbor magnetic moments' relative angles, enabled us to compute and categorize the ground state configurations and energies for any set of coupling parameters.

The complex phase diagram of the system was found to include a multitude of magnetic phases, such as ferromagnetic, antiferromagnetic, and non-collinear. Particularly, the non-collinear variants included a helical phase and two non-chiral phases, one of which exhibited ferromagnetic features. The non-collinear ferromagnetic phase exhibited a highly attenuated but persistent magnetization decay with increased system size. Additionally, we observed several magnetic phase transitions of both continuous and discontinuous types.

Dimensional reduction and clustering approaches (PCA and HDBSCAN) proved useful for distinguishing both collinear and non-collinear phases. However, the complexity of the system demanded an in-depth analysis and detailed modeling in order to attain a precise interpretation of the results.

Our work contributes to the understanding of magnetism in quasicrystalline systems and demonstrates the effectiveness of machine learning techniques for discovering new phases in complex systems. Future studies will focus on the characterization of magnetic excitations in the different phases and on the impact of temperature on the system's magnetic properties.

\begin{acknowledgments}
D.J.G. acknowledges support from PICT 2019-02396 of the ANPCyT. D.J.G. and M.N. acknowledge support from PIP 11220200101796CO of CONICET. P.S.C. acknowledges support from Grants PICT 2018-01546 and PICT 2019-00371 of the ANPCyT.
\end{acknowledgments}

% Path: bibliography.bib
%include bibliography files .bib
\bibliography{vdWmagnets,machinelearning,num_methods,fibo}

\appendix
\section{Energy minimization for classical magnetic moments} \label{ap:energymin}

We consider unit-length classical vectors to represent the magnetic moments. These vectors can be expressed as the product of a rotation matrix and a unit vector along the z-axis. The rotation matrix can be written as the exponential of a skew-symmetric matrix. Let $\vec{k}=(k_1,k_2,k_3)$ be a unit vector indicating the axis of rotation, and let $\theta$ be the angle of rotation. The skew-symmetric matrix can be written as:
\[
    S_k = \theta \begin{bmatrix} 0 & -k_3 & k_2 \\ k_3 & 0 & -k_1 \\ -k_2 & k_1 & 0 \end{bmatrix}
\]
The corresponding rotation matrix, $R$, can be obtained using Rodrigues' formula:
\[
R = e^{S_k} = I + \sin(\theta) S_k + (1 - \cos(\theta)) S_k^2
\]
Here, $R$ represents the rotation matrix derived from the exponential skew-symmetric matrix $S_k$.

Using these expressions we can represent the energy of the system as a function of the parameters $\{\theta k_1,\theta k_2,\theta k_3\}$ for each magnetic moment \cite{ivanov2021fast}. We can also calculate the derivative of the energy analytically.

With the expressions of the energy and of the derivative of the energy w.r.t. the different parameters that characterize the orientation of the magnetic moments, we can use the  Broyden-Fletcher-Goldfarb-Shanno (BFGS) method to minimize the energy.

We are using the version implemented in the Ensmallen library (flexible C++ library for efficient numerical optimization) \cite{curtin2021ensmallen}, which relies on the Armadillo library (a template-based C++ library for linear algebra) \cite{armadillocpp}.

\end{document}